\documentclass{elsarticle}
\biboptions{sort&compress}

\usepackage{color}
\usepackage{hyperref}
\usepackage{multirow}
\usepackage{graphicx}
\usepackage{amsmath}

\renewcommand{\vec}[1]{\textbf{\textit{#1}}}

\begin{document}

\title{Particle-in-Cell Laser-Plasma Simulation on Xeon Phi Coprocessors}
\author[unn]{I.A.~Surmin}
\ead{i.surmin@gmail.com}
\author[unn]{S.I.~Bastrakov}
\ead{bastrakov@vmk.unn.ru}
\author[unn,iap]{E.S.~Efimenko}
\ead{evgeny.efimenko@gmail.com}
\author[unn,iap,cut]{A.A.~Gonoskov}
\ead{arkady.gonoskov@gmail.com}
\author[unn,iap]{A.V.~Korzhimanov}
\ead{korzhimanov.artem@gmail.com }
\author[unn]{I.B.~Meyerov\corref{cor1}}
\ead{meerov@vmk.unn.ru}
\address[unn]{Lobachevsky State University of Nizhni Novgorod, Building 2, 23 Gagarina Avenue, Nizhni Novgorod, Russia 603950}
\address[iap]{Institute of Applied Physics of the Russian Academy of Sciences, 46 Ul'yanov Street, Nizhni Novgorod, Russia 603950}
\address[cut]{Chalmers University of Technology, SE-412 96, Gothenburg, Sweden}
\cortext[cor1]{Corresponding author}

\begin{abstract}

This paper concerns development of a high-performance implementation of the
Particle-in-Cell method for plasma simulation on Intel Xeon Phi coprocessors.
We discuss suitability of the method for Xeon Phi architecture and present our
experience of porting and optimization of the existing parallel
Particle-in-Cell code PICADOR. Direct porting with no code modification gives
performance on Xeon Phi close to 8-core CPU on a benchmark problem with 50
particles per cell. We demonstrate step-by-step application of optimization
techniques such as improving data locality, enhancing parallelization
efficiency and vectorization that leads to 3.75\,x speedup on CPU and 7.5\,x
on Xeon Phi. The optimized version achieves 18.8 ns per particle update on
Intel Xeon E5-2660 CPU and 9.3 ns per particle update on Intel Xeon Phi 5110P.
On a real problem of laser ion acceleration in targets with surface grating
that requires a large number of macroparticles per cell the speedup of Xeon
Phi compared to CPU is 1.6\,x.

\end{abstract}


\maketitle

\section{Introduction}

The progress in high intensity laser pulse generation throughout the last 20
years has stimulated theoretical and experimental research on ultra-intense
laser-matter interaction in extremely relativistic regimes \cite{Mourou,
Korzhimanov}. This is valuable for both fundamental research on physics of
matter in extreme conditions and various applications. The notable directions
are: laser-driven electron acceleration to the ultrarelativistic energies
\cite{Kostyukov}, acceleration of ion beams to tens and hundreds MeV / nucleon
\cite{Macchi, Bychenkov}, generation of X-ray and gamma ray radiation
including pulses of atto- and zeptosecond duration \cite{Teubner}, and QED
effects in ultrahigh intensity field, namely electron-positron pair production
and nonlinear optics of vacuum \cite{Di Piazza, Narozhny}. The applications
are fast ignition of inertial confinement fusion targets \cite{Ghoranneviss},
hadron therapy for cancer treatment \cite{Ledingham}, protonography, x-ray
imaging \cite{Albert}, etc.

High intensity laser-matter interaction involves several nonlinear physical
phenomena: relativistic and ponderomotive self-focusing, collisionless
heating, various plasma instabilities, high harmonics generation, and others.
The proper analytical examination is only possible in several simple cases.
Thus, one of the main tools for theoretical research is numerical simulation.
Along with investigation of the underlying physics, computer simulation helps
to set experiments by allowing faster and more efficient adjustment of
parameters for laser and target, designing experimental schemes, and
interpretation of the results.

The most widely used method for simulation of plasma in ultrahigh field is the
Particle-in-Cell (PIC) method \cite{Birdsal}, which allows to perform full
3D simulations that capture the main processes governing laser-plasma
interaction. The computational complexity of the Particle-in-Cell method is
relatively low compared to other kinetic methods (e.g. Euler \cite{Shoucri}), yet
large-scale 3D simulation requires high-performance implementation aimed at
supercomputers.
Currently, several widely known implementations of the fully relativistic
Particle-in-Cell method are capable of 3D large-scale plasma simulation on
supercomputers, most notably, OSIRIS \cite{Fonseca}, VPIC \cite{Bowers}, VLPL
\cite{Pukhov}, WARP \cite{Vay}, PIConGPU 
\cite{Burau}. The striking example of continuous progress in accelerating the
simulation in terms of both algorithms and implementation efficiency 
is a relativistic boosted frame that allows to speed up simulations of laser
propagation in low density plasma by a factor of tens \cite{Vay}.

Heterogeneous cluster systems are based not only on
convenient CPUs, but also on different types of accelerators, including
GPUs and the recently introduced Intel Xeon Phi coprocessors.
This results in a growing interest in Particle-in-Cell adaptations for
such systems. While the Particle-in-Cell method is generally suitable for
modern GPUs, its efficient implementation requires a meticulous approach to
data structures and parallel processing schemes as well as conscious usage of
different layers of GPU memory and is by no means a simple port of an
efficient implementation for CPUs.

Although several successful porting of applications to
Xeon Phi coprocessors have been reported
recently\cite{Jeffers_2014,Kulikov,Nakashima}, it is not certain if a
significant portion of peak performance is achievable for a wide class of
applications, and does efficient porting of an
existing application require code tuning in scope of OpenMP or rather massive
rewriting similar to porting to GPUs. A brief analysis shows that although a
straightforward porting can be done very quickly even for a large application,
it will be efficient only if the application was properly optimized for CPUs
and has a large degree of parallelism on thread-level and SIMD-level. This
might be a limiting factor, as many implementations scale well up to 8--16
threads, but not up to 120--240 threads or are only capable of using low width
SIMD. Another obstacle similar to GPUs might be a small amount of memory,
which on Xeon Phi is only 6--16 GB. Thus, a no-effort "just rebuild"
porting does not seem to be efficient except some very special cases, at the
same time a porting with reasonable additional tuning seems promising.

This paper presents our experience with porting the existing
parallel Particle-in-Cell plasma simulation code PICADOR
\cite{Bastrakov_2012,Bastrakov_2014} to Xeon Phi coprocessors. We demonstrate
a step-by-step optimization process with iterative bottleneck analysis and
application of optimization techniques. We believe that the encountered
problems and applied optimizations are not specific to the Particle-in-Cell
implementation and similar ideas are useful for a wide class of applications.
We illustrate porting and optimization of the code on CPUs and Xeon Phi coprocessors. We use a benchmark frozen plasma simulation problem with ideal balance and no MPI
exchanges. The performance of the final version is then evaluated on a real
problem of laser ion acceleration in targets with surface grating. This
problem is shown to be demanding a high number of macroparticles which makes
it highly suitable for acceleration by means of Xeon Phi coprocessors.

The paper is organized as follows. The Particle-in-Cell method is briefly
described in section \ref{sec_PIC}. In section \ref{sec_Phi} we discuss
suitability of the method for Xeon Phi coprocessors and
optimization methodology. We present experience of porting and optimization of
the Particle-in-Cell plasma simulation code PICADOR to Xeon Phi in section
\ref{sec_PICADOR}. Section \ref{sec_Real} contains performance evaluation of a
real simulation.

\section{Particle-in-Cell Method} \label{sec_PIC}

This section briefly describes the Particle-in-Cell method in the form used in
our implementation, a detailed description is given in \cite{Birdsal}.

The simulation area is a 3D axis-aligned parallelepiped covered by uniform
spacial grid. Dynamics of the electric field $\vec{E}$ and magnetic field
$\vec{B}$ is defined by the Maxwell's equations solved on the grid using the
FDTD method \cite{Taflove}. Plasma is represented as an ensemble of $N$
charged quasi-particles, each with a variable momentum $\vec{p}$ and position
$\vec{r}$, and constant mass $m$ and charge $q$. The position and velocity
$\vec{v}$ evolve according to Newton's law in relativistic form that is
numerically integrated using Boris method. Particles motion creates electric
current $\vec{j}$ that is a part of Maxwell's equations, enclosing the
self-consistent system of equations.

\begin{figure}
\noindent\centering{
\includegraphics[width=120mm]{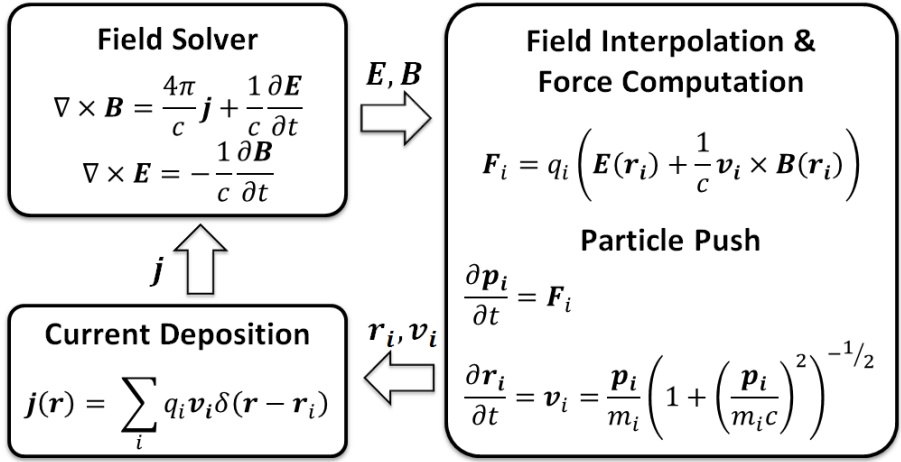}
}
\caption{Computational scheme of the Particle-in-Cell method. An iteration of the computational loop corresponds to a time step. The equations are given in the CGS system. Arrow labels denote data dependencies between stages.}
\label{fig1}
\end{figure}

A basic computational scheme of the Particle-in-Cell method with the main
equations and data dependencies is given in Fig.~\ref{fig1}. An iteration of
the computational loop corresponds to a time step. Each time step consists of
4 main stages. Field solver updates grid values of the electromagnetic field.
Field interpolation from the grid to particle position is performed to compute
the Lorenz force affecting particles. Solving equations of particle motion is
used to push particles one step further. The computational loop closes with
computation of current created by particle motion. In terms of software
implementation it is convenient to merge the field interpolation, force
computation and solving equations of particle motion into one stage, referred
as particle push. The Particle-in-Cell method can be extended in many ways
\cite{Gonoskov}, in this paper we consider only the basic version of the
method as these stages take significant share of computational time even in
more complicated simulations.

The Particle-in-Cell method operates on two principally different data sets:
an ensemble of charged particles with continuous coordinates and values of the
field and current density set on a discrete grid. The most time consuming
stages of particle push and current deposition operate on both data sets,
thus implementation of these stages plays a major role in both accuracy and
computational efficiency.

\section{Implementation of the Particle-in-Cell Method for Xeon Phi Coprocessors} \label{sec_Phi}

In this section we analyze suitability of the Particle-in-Cell method for Xeon
Phi coprocessors and discuss optimization techniques required to achieve good
performance. There are several highly efficient implementations of the method
for GPUs \cite{Burau} which implies good suitability for Xeon Phi as well.
However, there are several Xeon Phi-specific features and considerations.

\subsection{Overview of Xeon Phi Coprocessors}

A Xeon Phi coprocessor has 60 cores (61 on some modifications) on shared
memory with Linux on board. The cores are x86-compatible, but have simplified
architecture compared to CPU cores with no support to out-of-order execution
and branch prediction, combined with wider 512-bit vector registers (256-bit
on modern CPUs). Each core supports up to 4 hardware threads, it is
recommended to run at least 2 threads per core \cite{Jeffers_2013}. The peak
performance of a Xeon Phi core is lower than of a modern CPU core, but larger
number of cores provides overall nearly 1 TFLOPS peak performance in double
precision. Another notable advantage over CPUs is GDDR5 memory with about 350
GB/s peak throughput, 5 to 10 times larger compared to CPUs.

Programming for Xeon Phi can be done using traditional programming languages
and parallel programming libraries: C, C++, Fortran languages, MPI, OpenMP,
Cilk Plus, OpenCL, TBB and MKL libraries. This significantly simplifies
porting of existing applications as the vast majority of code (or even full
code) does not need to be modified for Xeon Phi. There are three execution
modes: running only on Xeon Phi as a multicore processor (native mode),
running the main program on a CPU and calling computational cores on a
coprocessor similar to GPU usage (offload), running some MPI processes on CPUs
and some on Xeon Phi coprocessors (symmetric) \cite{Jeffers_2013}.

Overall, Xeon Phi has manycore architecture with wide vector registers and
high throughput memory. Parallelism on both thread level (120--240 threads)
and 512-bit SIMD level is crucial for good performance. Although existing
applications can generally be ported without any significant effort,
performance of such ports is not guaranteed even for implementations that are
efficient on CPU. Some additional optimization and tuning is most likely
required. However, the main optimization principles for CPU and Xeon Phi
coprocessors are similar, so optimization for one of the platforms is likely
to yield benefit for the other.

\subsection{Analysis of Suitability of the Particle-in-Cell Method for Xeon Phi Coprocessors}
 
Generally there are several fundamental factors that determine applicability
of accelerators such as GPUs and Xeon Phi coprocessors. First, there is a
memory limitation of 6 to 16 GB RAM per accelerator which is coupled with a
PCI Express bus to the host. While this could be very restrictive for
applications working with large sets of data, Particle-in-Cell simulations are
spatially local, thus allowing to efficiently decompose a problem between many
computational nodes and fit memory limitations of a node.

Performance-wise Xeon Phi coprocessors offer about 5\,x peak performance
advantage over top server CPUs. Achieving a significant share of peak
performance is challenging on CPUs and even more so on Xeon Phi coprocessors.
However, a half of peak performance of Xeon Phi is attributed to vector fused
multiply-add (FMA) instruction that performs two floating-point operations at
once. Obviously, real applications do not consist of sequences of pure FMA
calls and not utilizing FMA effectively reduces performance by half.

Efficient utilization of Xeon Phi requires excellent scaling on shared memory.
Coprocessors offer 60 cores with up to 4 hardware threads. Well-suited
applications usually either scale well up to 120--240 threads or use efficient
hybrid MPI + OpenMP parallel scheme. One way or another, the method must have
a significant parallelization potential. An ideal application in this regard
(for pretty much all parallel hardware) is a Monte-Carlo method, which allows
ideal scaling. The Particle-in-Cell method also allows ideal scaling
on shared memory CPU cores and, in our experience, 75\% scaling efficiency on
Xeon Phi cores (for large number of particles per cell). It is important to
note that good scalability is only the necessary condition for efficient
utilization of Xeon Phi, but not sufficient.

Another important --- and often the most challenging --- factor is utilization
of vector units for performing floating-point operations, so-called
vectorization. Vectorization is important for CPUs, and even more so on Xeon
Phi due to 512-bit vector registers compared to 256-bit on modern CPUs. As
consequently, vectorization theoretically offers double dividend on Xeon Phi
compared to CPUs and lack of vectorization puts double penalty on Xeon Phi
performance. The Particle-in-Cell method is not an easy candidate for
vectorization. A straightforward implementation of field interpolation and
current deposition in 3D results in non-unit stride (not local) memory access
pattern that is detrimental to vectorization. A special organization of those
operations is required to allow proper vectorization.  

Our analysis shows that the Particle-in-Cell method, although not ideally
suited for Xeon Phi coprocessors due to vectorization issues, is a promising
candidate. We will show that with proper programming a Xeon Phi coprocessor
can achieve about 2\,x speedup over a tuned implementation on a modern 8-core
CPU on both benchmark and simulation of the real problem.

\subsection{Optimization Methodology}

Although some computational applications are developed for specific class of
hardware and thus could largely benefit from hardware-specific features, the
vast majority of applications are initially developed for CPUs and then ported
to accelerators. This subsection is devoted to general discussion of porting
and optimization of an existing computational application from CPUs to
accelerators. We suppose the numerical schemes being used are suited for
parallel processing and the code is developed using standard for HPC
programming languages C, C++ or Fortran and parallel programming technologies
MPI and OpenMP.

By all means, one should start with optimization for modern CPUs mainly
focusing on scaling efficiency and vectorization. In most cases this
optimization is beneficial for Xeon Phi as well. An existing code can be
ported to Xeon Phi by just recompiling it with Intel Compiler; this is a
substantial advantage of Xeon Phi over GPUs that generally require significant
modification of the code. In our experience this "no-effort" port usually
requires several minutes in an ideal case to several hours, mostly spent on
rebuilding third-party libraries. Performance of such port can be discouraging
even for codes that are fairly efficient on CPUs. It is essential to use
profiling tools to discover the most time-consuming routines and
performance-limiting factors. Pieces of code that were fast enough to not
optimize and parallelize on CPU often become a bottleneck on Xeon Phi due to
larger amount of cores and poor single-core performance. After such unexpected
bottlenecks are eliminated, one should generally focus on scaling efficiency
and vectorization in the most time-consuming routines and loops. 

We have employed this approach to porting our implementation of the
Particle-in-Cell method, gradually profiling, solving performance issues and
measuring effect of optimization on CPU and Xeon Phi. Only after our
optimization resources on CPU were exhausted, we proceeded to trying Xeon
Phi-specific optimizations. Those included manual vectorization using
intrinsics translated into vector instructions of Xeon Phi, using large memory
pages to reduce DTLB miss rate, non-temporal stores for better cache
utilization efficiency, adjusting number of processors and threads in MPI +
OpenMP scheme. Most of the enlisted optimization techniques did not yield a
significant if any performance benefit for our application, a notable
exception is manual vectorization via intrinsics.

\section{Porting and Optimization of the Particle-in-Cell Code PICADOR} 
\label{sec_PICADOR}

\subsection{PICADOR Particle-in-Cell Code }

PICADOR \cite{Bastrakov_2012, Bastrakov_2014} is a fully parallel 3D
Particle-in-Cell implementation capable of running on heterogeneous cluster
systems with CPUs, GPUs and Xeon Phi coprocessors. Features of PICADOR include
FDTD and NDF field solvers, Boris particle pusher, CIC and TSC particle form
factors, Villasenor-Buneman and Esirkepov current deposition, ionization,
moving frame, and dynamic load balancing. Each MPI process handles a part of
simulation area (domain) using either a multicore CPU or Xeon Phi coprocessor via
OpenMP, or GPU via CUDA. All MPI exchanges occur only between
processes handling neighboring domains.

The baseline version was developed taking into account some common knowledge
performance considerations, the short description is given below. 

Values of each vector component are stored separately in a 3D array wrapped
into 1D. Current values are written directly to global current arrays,
similarly, field values are read directly from global field arrays. The key performance consideration is a particle storage. We use separate array of particles per each cell with Array-of-Structures (AoS) layout. This approach varies from the widely used global particle array with sorting strategy and is closer to data structures used for GPU-based
implementation (but without supercells common for GPUs).

On the most time consuming particle push and current deposition stages
particles are processed in a cell-by-cell order. Particles in several cells
are processed in parallel using OpenMP in one-pragma style. After each
particle push we perform a check and, in case a particle leaves the current
cell, update particle storage structure accordingly. This migration check is
done partly in parallel: each thread has its own buffer for migrating
particles, after all checks are done all buffers are merged and processed
sequentially.  On the current deposition stage each thread writes to its own
current buffer to avoid data races, all buffers are summed after the stage is
over.

\subsection{Benchmark, Hardware and Performance Measurement Details}

For all the performance measurements presented in this paper we used a frozen
plasma benchmark with $40 \times 40 \times 40$ grid, 50 particles per cell,
and 1000 time steps. We used CIC particle formfactor for field interpolation
and current deposition, and perform all calculations in double
precision. The time given in tables refers only to the computational phase
which is a sum of particle push, current deposition and field update.

Computational experiments were done on a node of Lobachevsly cluster system at
University of Nizhni Novgorod with 8-core Intel Sandy Bridge E5-2660 CPUs (2.2
GHz), 64 GB RAM, and 2 Intel Xeon Phi 5110P coprocessors, each with 60 cores,
240 threads, and 8 GB RAM. Peak performance of each CPU in double precision is
140 GFLOPS, and peak performance of Intel Xeon Phi 5110P is 1 TFLOPS. The code was compiled with Intel C++ Compiler.

Table \ref{TableBaseline} presents performance results of the baseline version on the CPU and Xeon Phi. On Xeon Phi we used native mode and porting required only rebuilding the code and libraries with compiler options for Xeon Phi support (-mmic). Performance on Xeon Phi is very close to CPU, but time distribution between stages is different with faster particle push and significantly slower field update. Thus, no effort port gives reasonable performance (taking in account in only took several hours) but further optimization is required.
\begin{table}[h!]
\caption{Performance of the baseline version on CPU and Xeon Phi}
\label{TableBaseline}
\begin{center}
\begin{tabular} {ccc}
\hline
\multirow{ 2}{*}{\textbf{Stage}} & \multicolumn{2}{c}{\textbf{Time [s]}} \\ {} & \textbf{CPU} & \textbf{Xeon Phi} \\
\hline
Particle push & 163.1 & 134.8\\
Current deposition &  61.3 & 81.3 \\
Field update &  0.8 & 7.7 \\
Total & 225.2 & 222.8 \\ 
\hline
\end{tabular}
\end{center}
\end{table}

\subsection{Improving Memory Locality}

A natural and widely used idea for efficient implementation of the
Particle-in-Cell method is to use physical locality of the method --- each
particle is interacting only with several closest field values --- and
transform it into memory locality to allow cache-friendly implementation. With
CIC particle formfactor and Yee grid particles in each cell interact only with
27 closest grid values for each field and current component (a cube with side
3). Before processing particles of a cell we preload corresponding 27
surrounding field values into a small local array and use these values for
field interpolation. In a similar way, we accumulate currents created by
particles of a cell in a small local array and add it to the global array
after all particles are processed. Thus, we replace the majority of memory
operations with global field and current arrays with the same operations on
local arrays.

The comparison of performance of this version and the baseline version is
presented in Table \ref{TableMemory}. Improving memory locality yields over
3\,x benefit over the baseline on both CPU and Xeon Phi.
\begin{table}[h!]
\caption{Performance of the version with improved memory locality}
\label{TableMemory}
\begin{center}
\begin{tabular} {ccccc}
\hline
\multirow{ 2}{*}{\textbf{Stage}} & \multicolumn{2}{c}{\textbf{Time [s]}} & \multicolumn{2}{c}{\textbf{Speedup to the baseline}} \\ {} & \textbf{CPU} & \textbf{Xeon Phi} & \textbf{CPU} & \textbf{Xeon Phi}\\
\hline
Particle push &  56.9 & 41.3 & 2.87\,x &  3.26\,x \\
Current deposition&  14.0 & 16.9 & 4.38\,x & 4.81\,x \\
Field update &  0.8 & 7.7 & 1.00\,x & 1.00\,x \\
Total & 71.7 & 65.9 & 3.14\,x & 3.38\,x \\ 
\hline
\end{tabular}
\end{center}
\end{table}

\subsection{Enhancing Scalability on Shared Memory}

Efficiency of scaling on shared memory is important for multicore CPUs and
even more for Xeon Phi. First, we changed parallel current deposition
scheme. While storing a separate global current array for each thread to avoid
data races is possible for 16 threads it is probably not the most efficient
way and definitely not practical for 240 threads of Xeon Phi. Thus, we
developed a new parallel current deposition scheme that does not replicate
global current array. Again, we employ locality properties described in the
previous subsection. Since particles in each cell only contribute to grid
values in $3 \times 3 \times 3$ surrounding cube, particles that are distant
enough from one another can be processed in parallel without any risk of data
races. Namely, for CIC form factor we can concurrently process cells that have
2 unprocessed cells in between; the same idea can be applied to other form
factors with probably larger distance. Thus, current deposition consists of 27
particle traversals in checkerboard order, each internal traversal is parallel
with only synchronization at the end of the traversal.

Then we eliminated sequential migration of particles between cells. Due to
relation between space and time steps, particle can not pass the distance
greater than cell size in one time step. Thus, a migrating particle is
necessarily located in a neighbor cell. We again applied the checkerboard
order parallelization scheme. For each cell we create a buffer for particles
migrating to this cell, after pushing a particle each thread computes new cell
index and in case of migration writes the particle to a buffer of the new
cell. Because of locality and checkerboard traversal order there is not need
for synchronization except at the end of each traversal.

The performance of this version is presented in Table \ref{TableScaling}.
\begin{table}[h!]
\caption{Performance of the version with improved memory locality and enhanced scalability}
\label{TableScaling}
\begin{center}
\begin{tabular}{ccccc}
\hline
\multirow{ 2}{*}{\textbf{Stage}} & \multicolumn{2}{c}{\textbf{Time [s]}} & \multicolumn{2}{c}{\textbf{Speedup to the baseline}} \\ {} & \textbf{CPU} & \textbf{Xeon Phi} & \textbf{CPU} & \textbf{Xeon Phi}\\
\hline
Particle push & 52.1 & 37.1 & 3.13\,x & 3.63\,x \\
Current deposition & 13.9 & 13.1 &  4.41\,x & 6.21\,x \\
Field update &  0.7 & 1.8 & 1.14\,x & 4.28\,x \\
Total & 66.7 & 52.0 & 3.38\,x & 4.28\,x \\ 
\hline
\end{tabular}
\end{center}
\end{table}

\subsection{Improving Vectorization}

Efficient vectorization is a key factor in achieving good performance on CPUs
and particularly on Xeon Phi coprocessors. The main reason for lackluster
performance on Xeon Phi is poor vectorization of the code.

We tried two approaches to vectorization. First we tried to assist the complier with auto-vectorization by using special directives of Intel Compiler (such as \#pragma ivdep and \#pragma simd) and
loop splitting. It allowed to vectorize implementation of field update and the
Boris method for particle push, but did not vectorize more time-consuming
field interpolation and current deposition because of complicated memory
access pattern. Thus, vectorization lead to a modest speedup of particle push
and field update, as shown at Table \ref{TableVectorization}, this version for
Xeon Phi is denoted as v1.

\begin{table}[h!]
\caption{Performance on CPU and Xeon Phi after improving vectorization. Compiler auto-vectorization version on Xeon Phi is denoted as v1, version with manual vectorization of field interpolation and current deposition is denoted as v2.}
\label{TableVectorization}
\begin{center}
\begin{tabular}{ccccc}
\hline
\multirow{ 2}{*}{\textbf{Stage}} & \multicolumn{2}{c}{\textbf{Time [s]}} & \multicolumn{2}{c}{\textbf{Speedup to the baseline}} \\ {} & \textbf{CPU} & \textbf{Xeon Phi v1/v2} & \textbf{CPU} & \textbf{Xeon Phi v1/v2} \\
\hline
Particle push & 45.6 &  20.2 / 18.8 & 3.58\,x & 6.67\,x / 7.17\,x \\
Current deposition & 13.8 & 13.0 / 10.1 & 4.44\,x & 6.25\,x / 8.05\,x \\
Field update & 0.6 & 0.8 / 0.8 & 1.33\,x & 9.63\,x / 9.63\,x \\
Total & 60.0 & 34.0 / 29.7 & 3.75\,x & 6.55\,x / 7.50\,x \\ 
\hline
\end{tabular}
\end{center}
\end{table}

The main reason of lackluster speedup due to vectorization is that for each
cell we have to store $3 \times 3 \times 3$ arrays of field and current
density components, while each particle uses $2 \times 2 \times 2$ subarray
depending on its position inside the cell. The implementation uses indirect
indexing which renders vectorization of the loop over particles inefficient on
Xeon Phi.  

To eliminate indirect indexing the field and current density values have been
repacked into eight $2 \times 2 \times 2$ arrays each corresponding to an octant
of a cell. Those arrays contain 512-bit vector elements of the components of
the electric and magnetic field (6 values) for field interpolation and the
components of the current density (3 values) for current deposition. Thus,
field interpolation effectively uses 75\% of the vector register length and
current deposition uses only 37.5\%. For each particle we determine the octant
it belongs and perform the corresponding operation with vector registers using
intrinsics. This modification was only done for Xeon Phi as vector extensions
are different for Xeon Phi and CPUs and yield additional 1.15\,x overall
speedup, as shown at Table \ref{TableVectorization}.

The main obstacle for efficient vectorization is indirect addressing caused by
usage of the standard Yee grid for values of the electro-magnetic field and
current density \cite{Taflove}. As shown in \cite{Nakashima}, vectorization
can be done much easier for a more straightforward grid.

\section{Performance Evaluation on a Real Simulation} \label{sec_Real}

\subsection{Problem Statement}
In this section we present the results of efficiency
measurements of our Xeon Phi implementation solving a real physical problem.
As an example, we chose a laser ion acceleration in a so-called Target Normal
Sheath Acceleration (TNSA) regime \cite{Wilks}. In this regime ions 
are accelerated from a rear side of a thin (sub-micron thick) solid-state
target by quasistatic electron sheath. This sheath is formed by hot electrons 
accelerated to relativistic energies by laser field at front surface of the
target. One of the main problems of the scheme is its low efficiency 
\cite{Bychenkov}. Recently, it has been suggested to use surface grating on
the irradiated size of the target to increase laser-electron coupling and
therefore increase amount of energy transferred from laser radiation to ions
\cite{Nodera}. This proposal has been tested via a number of 2D
numerical simulations \cite{Takahashi, Pae, Andreev}, however for more
realistic results full 3D simulations are needed especially for investigation
of complex gratings. Realistic 3D simulations, however, are known to be
extremely resource-demanding and thus may greatly benefit from the opportunity
to be run on heterogeneous systems enabling resources of Xeon Phi
coprocessors.

We investigated irradiation of a 0.3~$\mu$m thick target composed of
Au$_{197}^{+31}$ ions with plasma density corresponding to electron
concentration $3\times 10^{21}$~cm$^{-3}$ (overdense parameter $n_0 = 30$).
The accelerated proton layer with the same electron concentration and
0.1~$\mu$m thickness has been attached to a rear side of the target. The
grating in the form of rectangular brush modulated along y-axis only has been
placed on an irradiated side. The grating height has been chosen to be equal
to 0.3~$\mu$m, the thickness of a single element was equal to 0.15~$\mu$m and
the grating period was equal to 0.5~$\mu$m. Initial temperature of all plasma
components was 100~eV which is referred to values usually expected from
collisional heating.

The laser pulse normally impinged on the target was supposed to be an infinite
in transversal direction plane wave propagating along x-axis and linearly
polarized in the y-direction. It had a Gaussian envelope along propagation
axis with duration at full width half maximum equal to 42~fs and its
wavelength was equal to 1~$\mu$m. The laser intensity in maximum reached
$3.75\times 10^{19}$~W/cm$^2$ (dimensionless amplitude $a_0=5.2$). These
parameters are typical for Ti:Sa terawatt systems widespread nowadays in
laboratories worldwide.

The simulation area is a box of size $12\times 1\times 1$~$\mu$m covered with
$512\times 64\times 64$ grid. A time step was equal to $0.026$~fs and the
total simulation time was $300$~fs, requiring 11\,512 time steps to complete.

\begin{figure}
\noindent\centering{
\includegraphics[width=100mm]{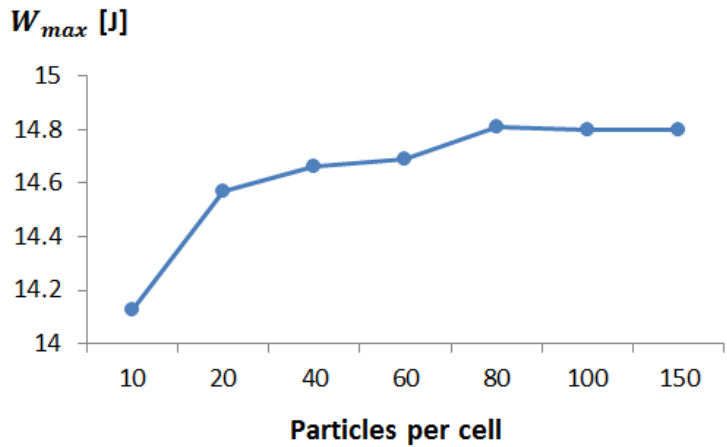}
}
\caption{Dependence of the maximal ion energy on the number of particles per cell.}
\label{max_ion_en_vs_ppc}
\end{figure}

The maximal ion energy $W_{max}$ is sensitive to the number of macroparticles
per cell $N_{PPC}$. This is due to the fact that higher $N_{PPC}$ provides
better resolution of a tail of sheath electron energy distribution and this
tail is known to define the maximal energy $W_{max}$ reached by the
accelerated ions. In our simulations we varied PPC-parameter in 10 to 150
range in order to investigate the dependence of $W_{max}$ on $N_{PPC}$,
which is shown in Fig.~\ref{max_ion_en_vs_ppc}. One can see that obtained
maximal ion energy steadily grows with increasing $N_{PPC}$ until it reaches
14.8~MeV at $N_{PPC} = 80$ and does not grow further. This feature of the
investigated problem makes it highly favourable for running on Xeon Phi
coprocessors because in comparison with CPUs they do Particle-in-Cell
simulations with high $N_{PPC}$ faster.

\subsection{Performance Evaluation}

\begin{figure}
\noindent\centering{
\includegraphics[width=120mm]{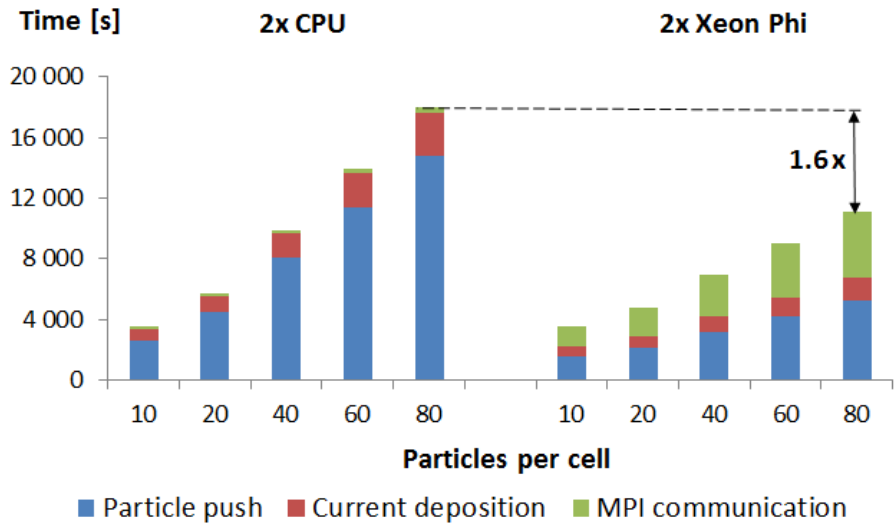}
}
\caption{Performance of 2x CPUs and 2x Xeon Phi coprocessors on a TNSA simulation with 10 to 80 macroparticles per cell.}
\label{fig3}
\end{figure}

We measured performance of simulations on 2 CPUs and 2 Xeon Phi coprocessors
with the number of particles per cell varying from 10 to 80. The results are
presented in Fig.~\ref{fig3}. Three most time-consuming stages are particle
push, current deposition and MPI exchanges. Other stages, including field
solver, pulse generation and absorbing boundary conditions took negligible
time.

MPI communication phase on Xeon Phi takes much longer not only due to
additional data transfer between coprocessor and host occurring with each MPI
data exchange, but mainly due to preparation of the data to be transfered
involving copying of the boundary of 3D arrays to 1D arrays, that is not fully
parallelized. Total computational time on CPU and Xeon Phi is close for
$N_{PPC} = 10$ with Xeon Phi steadily outperforming CPU as $N_{PPC}$ grows.
For $N_{PPC} = 80$ simulation on Xeon Phi is 1.6\,x times faster compared to
CPU with 2.8\,x speedup of particle push, 2.0\,x speedup of current
deposition, 2.6\,x overall speedup of computational core, and significantly
slower data exchanges and related operations. 

\section{Summary}

This paper studies suitability of Intel Xeon Phi coprocessors as accelerators
for Particle-in-Cell plasma simulation. We discuss features of Xeon Phi that
influence performance of Particle-in-Cell implementations. Limitation of 16 GB
RAM on Xeon Phi is not a severe constraint for Particle-in-Cell simulation.
Due to massive parallelism potential of the Particle-in-Cell method it can
efficiently utilize Xeon Phi.  We found that an important limiting factor is
low single-core performance, particularly for code that is not ideally
vectorized. In practice, it means that the pieces of code fast enough
to not be parallelized or vectorized on CPU can become a bottleneck on Xeon
Phi and therefore has to be rewritten.  

We confirm this conclusions by porting of an existing 3D
Particle-in-Cell plasma simulation code PICADOR to Xeon Phi coprocessors. The
parallel C++ code using MPI and OpenMP was originally ported by means of just
recompiling with performance on Xeon Phi close to that of 8-core CPU. We
demonstrate step-by-step application of standard optimization techniques:
improving memory locality, scaling efficiency and vectorization that lead to
overall speedup of 3.75\,x on CPU and 7.5\,x on Xeon Phi. On a test problem
with 50 macroparticles per cell the final version achieves 18.8 ns per
particle update on CPU and 9.3 ns per particle update on Xeon Phi in double
precision. On the real simulation of laser ion acceleration two Xeon Phi
coprocessors outperform two CPUs by factor of 1.6. Our implementation
demonstrates good scaling on shared memory with 240 threads of Xeon Phi. The
most challenging aspect of efficient implementation is vectorization of field
interpolation and vectorization due to non-unit stride memory access. Overall,
the Particle-in-Cell method, although not ideally suited for Xeon Phi
coprocessors due to vectorization issues, allows to achieve significant
performance gain on a modern heterogenous clusters.

In terms of further development, the second-generation Intel
Knights Landing coprocessors has significant advances over the current
Xeon Phi architecture including significant improvement of the single-core
performance and support for out-of-order execution, and capability for direct
data exchanges between coprocessors avoiding PCI Express. Higher single-core
performance will allow efficient Particle-in-Cell simulation with lower number
of macroparticles per cell and to a certain extent alleviate performance
degradation of a poorly vectorized or parallelized code. The capability for
coprocessor-only simulation with direct data transfers provides good scaling
potential. This improvements together with our good experience of porting
PICADOR to existing hardware makes Xeon Phi coprocessors a very promising
hardware platform for high-performance Particle-in-Cell plasma simulation.

This study was partially supported by the RFBR, research project No.~14-07-31211 and by the grant (the agreement of August 27, 2013 No. 02.Â.49.21.0003 between The Ministry of education and science of the Russian Federation and Lobachevsky State University of Nizhni Novgorod).

\end{document}